\documentclass[]{aa}  

\usepackage{graphicx}
\usepackage{txfonts}
\usepackage{xcolor}
\usepackage{ulem}

\newcommand{\kmsmpc}{$\textrm{km} \ \textrm{s}^{-1} \textrm{Mpc}^{-1}$}
\newcommand{\kms}{$\textrm{km} \ \textrm{s}^{-1}$}
\usepackage{hyperref}
\hypersetup{
    colorlinks=true,
    urlcolor=cyan,
    citecolor=blue,
}
\begin{document}

   \title{J1721$+$8842: a gravitationally lensed binary quasar with a proximate damped Lyman-$\alpha$ absorber}


   \author{C. Lemon
          \inst{\ref{epfl}}, M. Millon\inst{\ref{epfl}},  D. Sluse\inst{\ref{liege}}, F. Courbin\inst{\ref{epfl}}, M. Auger\inst{\ref{camb}, \ref{camb2}}, J. H. H. Chan\inst{\ref{epfl}}, E. Paic\inst{\ref{epfl}} \and A. Agnello \inst{\ref{dark}}
          }
    
\authorrunning{C. Lemon et al.}
\institute{
Institute of Physics, Laboratory of Astrophysics, Ecole Polytechnique F\'ed\'erale de Lausanne (EPFL), Observatoire de Sauverny, 1290 Versoix, Switzerland \label{epfl}
\and
STAR Institute, Quartier Agora - Alle du six Aout, 19c B-4000 Liege, Belgium \label{liege}
\and
Institute of Astronomy, University of Cambridge, Madingley Road, Cambridge CB3 0HA, UK \label{camb}
\and
Kavli Institute for Cosmology, University of Cambridge, Madingley Road, Cambridge CB3 0HA, UK \label{camb2}
\and
DARK, Niels Bohr Institute, Jagtvej 128, 2200 Copenhagen, Denmark \label{dark}}

   \date{August 25, 2021}

 
\abstract{
High-redshift binary quasars provide key insights into mergers and quasar activity, and are useful tools for probing the spatial kinematics and chemistry of galaxies along the line-of-sight. However, only three sub-10-kpc binaries have been confirmed above $z=1$. Gravitational lensing would provide a way to easily resolve such binaries, study them in higher resolution, and provide more sightlines, though the required alignment with a massive foreground galaxy is rare. Through image deconvolution of StanCam Nordic Optical Telescope (NOT) monitoring data, we reveal two further point sources in the known, $z \approx 2.38$, quadruply lensed quasar (quad), J1721+8842. An ALFOSC/NOT long-slit spectrum shows that the brighter of these two sources is a quasar with $z = 2.369 \pm 0.007$ based on the \textsc{C iii]} line, while the \textsc{C iii]} redshift of the quad is $z = 2.364 \pm 0.003$. Lens modelling using point source positions rules out a single source model, favouring an isothermal lens mass profile with two quasar sources separated by $\sim6.0$ kpc (0.73\arcsec) in projection. Given the resolving ability from lensing and current lensed quasar statistics, this discovery suggests a large population of undiscovered, unlensed sub-10-kpc binaries. We also analyse spectra of two images of the quad, showing narrow \textrm{Ly$\alpha$} emission within the trough of a proximate damped Lyman-$\alpha$ absorber (PDLA). An apparent mismatch between the continuum and narrow line flux ratios provides a new potential tool for simultaneously studying microlensing and the quasar host galaxy. Signs of the PDLA are also seen in the second source, however a deeper spectrum is still required to confirm this. Thanks to the multiple lines-of-sight from lensing and two quasar sources, this system offers simultaneous sub-parsec and kpc-scale probes of a PDLA. 
  }


   \maketitle
%

\section{Introduction}
Quasar clustering is a natural prediction of the hierarchical structure formation scenario, and the distributions of luminosity, redshift, and separation, shed light on mergers and their role in the onset of quasar activity \citep{djorgovski1991, hennawi2006, hopkins2007, daangela2008}. Thanks to dedicated spectroscopic confirmation campaigns in recent years, the clustering amplitude is well-measured above $\sim$25\,kpc up to $z\approx$ 2, i.e., sky separations above 3 arcseconds \citep{hennawi2010, kayo2012, eftekharzadeh2017}. However, probing the final stages of mergers at the the peak of quasar activity ($z\sim$ 2$-$3) requires yet smaller separation binaries at higher redshifts, of which very few are known (see \citealt{chen2021} for a compilation of redshift and separations of known systems and candidates). Both observations and simulations link increased quasar activity with decreasing merger separation, peaking below 10 kpc, suggesting a similar peak should be seen for binary quasars \citep{vanwassenhove2012, capelo2017, stemo2020}. Whether such systems are simply intrinsically rare due to astrophysical processes, such as the restriction of gas inflow as mergers become stable \citep{mortlock1999}, or are simply undiscovered, is not yet known \citep{foreman2009}. Thanks to high-resolution wide-field surveys like \textit{Gaia} and HSC, targeted searches for such systems is becoming possible, but requires extensive spectroscopic and imaging follow-up \citep{shen2021, tang2021}. The handful of confirmed high-redshift quasar pairs at $\sim$1$\arcsec$ separation are known thanks to large spectroscopic follow-up campaigns of gravitational lens candidates \citep{more2016, schechter2017, anguita2018, lemon2018}. 

Gravitational lenses themselves provide important tools for probing galaxy formation and evolution, and also cosmology. The positions and fluxes of the multiple images constrain the lensing galaxy mass, and can test Cold Dark Matter through sub-halo number measurements \citep{gilman2019, hsueh2020}. The magnification accompanied by lensing allows us to study faint high-redshift source populations at high-resolution \citep{paraficz2018, hartley2019, inoue2020, ding2021, stacey2021}. Lensed quasars in particular offer the unique probe of time-delay cosmography, due to their intrinsic source variability \citep{liao2019, millon2020a, millon2020b, harvey2020, wong2020, shajib2020}. Furthermore, microlensing by stars in the lensing galaxy can be used to study quasar structure \citep{sluse2015, hutsemekers2020, cornachione2020, shalyapin2021}, and the Initial Mass Function and primordial black holes \citep{mediavilla2017, jimenezvicente2019, estebangutierrez2020, hawkins2020}. Given the bright quasar source and multiple lines of sight, characterising the metallicity, geometry, and structure of intervening material provides strong observational constraints on the anisotropy and densities of elements in galaxies out to high redshift \citep{smette1992, okoshi2019, cashman2021}.

The search for gravitationally lensed quasars is fuelled by the lack of known systems for all of these science cases, and further hindered by the need for specific characteristics adapted to the particular probe (e.g., asymmetric quads for time delays, broad-absorption line quasar sources for certain source probe techniques, etc.). The ongoing discovery and characterisation of these objects is needed to advance such studies. 

In this paper, we describe the initial results from targeted imaging and spectroscopy of a recently discovered lensed quasar, J1721+8842. The lens was discovered by \citet{lemon2018} by searching for multiple \textit{Gaia} detections around quasar candidates with quasar-like WISE colours, and subsequently graded with Pan-STARRS imaging. Basic pixel modelling showed a residual next to one of the four quasar images, which the authors suggest to be either a foreground object or a possible further image of the source. 

In Sect. \ref{imaging}, we describe the imaging and deconvolution from monitoring data at the Nordic Optical Telescope (NOT). In Sect. \ref{notspectrum}, we analyse a long-slit spectrum from ALFOSC on the NOT, followed by a targeted spectrum for the lens redshift in Sect. \ref{whtspectrum}. In Sect. \ref{modelling}, we perform mass modelling to infer the likely number of sources, and discuss the results in Sect. \ref{discussion}. We conclude in Sect. \ref{conclusions}. Throughout we use a flat-Lambda Cold Dark Matter cosmology with $H_{0}$ = 70 \kmsmpc, $\Omega_{M}$ = 0.3, and $\Omega_{\Lambda}$ = 0.7.

\section{Image Deconvolution} \label{imaging}
The proximity of J1721+8842 to the north celestial pole ideally provides no season gaps for a possible time-delay measurement. However, it is either inaccessible or at perpetually high airmass to many northern telescope facilities. Its bright and well-separated images overcomes this drawback and so J1721+8842 was targeted for photometric monitoring with StanCam (pixel scale of 0.176\arcsec\ pixel$^{-1}$) on the NOT under Programs 63-804 (P.I.: F. Courbin) and 63-501 (P.I.: A. Agnello). Five exposures of 280s are taken each night. The data used in this paper are from 3 April 2021 -- 2 July 2021.

With the eventual goal of measuring time delays from these monitoring data, single-epoch photometry must be extracted from seeing-limited observations. Disentangling the lensing galaxy and lensed source host galaxy light from the point sources is required for reliable photometry. One well-tested technique has been the MCS deconvolution \citep{mcs}, and has been used to obtain millimag-precision lightcurves for many lensed quasars \citep[see, e.g.,][]{millon2020a, millon2020b}. The process is fully described in \citet{cantale2016}. In brief, each image is deconvolved with a model of the PSF that is not the observed PSF but a narrower one, ensuring that the resolution in the deconvolved image is consistent with its sampling. The deconvolved image can have an arbitrary PSF shape, which we choose as a circular Gaussian with 2 pixels FWHM. In addition, the sampling adopted in the deconvolved image can be as fine as desired. A common choice is to adopt an oversampling of 2 with respect to the original data. Finally, the deconvolved image is decomposed into two separate channels, one containing all (Gaussian) point sources (here the quasar images) and one containing a free-form representation of any extended object in the data (here the lensing galaxy and the quasar host). All images in the monitoring time series are deconvolved jointly, with each image having its own PSF. In doing so, the intensities of the PSFs in each frame are free parameters, and the deconvolved extended-channel is an array of pixels to which regularisation is applied. The extended channel is shared by all images in the time series, as well as the positions of the point sources. The final output of the process is a deep, sharp deconvolved image with fine sampling which has the signal-to-noise of the whole dataset and the intensities of all point sources at all monitoring epochs. Before applying this technique, the frames are bias-subtracted and flat-fielded, following a sky background subtraction using SExtactor \citep{bertin1996}.

\begin{figure}
\centering
\includegraphics[width=0.5\textwidth]{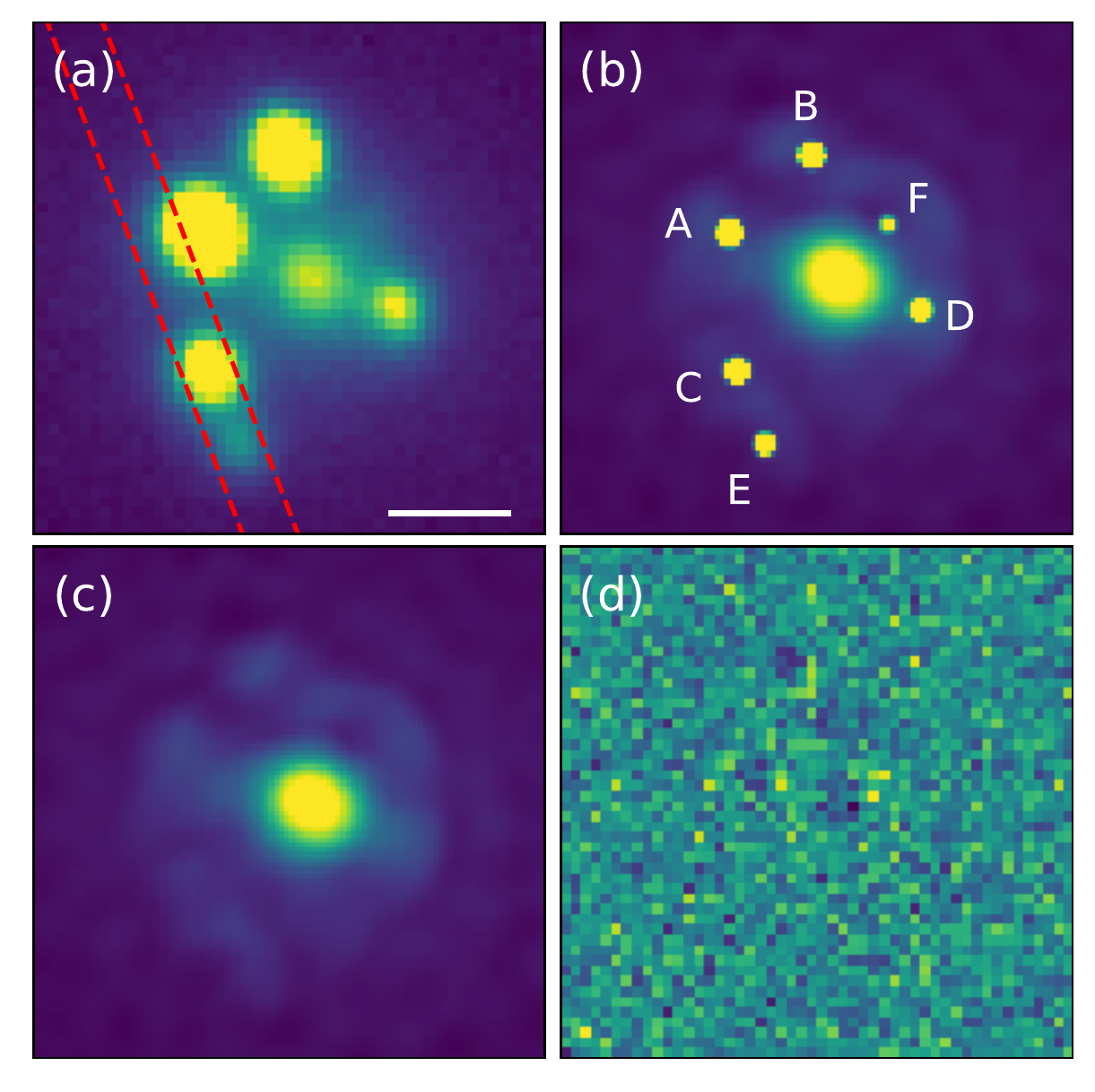}
\caption{StanCam imaging and deconvolution of J1721+8842, with North up, East left. (a) Best-seeing stack image using 34 epochs with seeing < 1.1\arcsec, with the ALFOSC slit overlaid; (b) deconvolved image using all epochs; (c) deconvolved image removing the point source channel; (d) residuals after subtracting the best-fit model from the best-seeing single-epoch image. The white scale bar in (a) is 2\arcsec.}
          \label{deconvolution}%
\end{figure}

The deconvolution requires iteratively correcting the positions of known point sources. Original deconvolutions with just 4 PSFs left two compact residuals either side of the lensing galaxy. To obtain clean residuals across all epochs, two further point sources were added, producing the results shown in Fig. \ref{deconvolution}. The second and early-third \textit{Gaia} data release catalogues contain five detections for this system, corresponding to the quad images and E, as labelled in Fig. \ref{deconvolution} \citep{gaia2, gaia3}. The astrometric excess noise, a catalogue parameter often used to assess compatibility with an isolated point source \citep[e.g., ][]{belokurov2017}, is zero and thus consistent with a PSF for E. The fainter of these two point sources, F, is not detected in these catalogues, as its flux is below the detection limit. The best-fit astrometry from the StanCam imaging (measured simultaneously on all epochs) and best-seeing frame flux ratios are listed in Table \ref{table:astrometry}. The lensing galaxy position and uncertainties are found by fitting a Sersic profile convolved by a 2-pixel FWHM Gaussian to the deconvolved background light (lower right panel of Fig. \ref{deconvolution}), while the uncertainties on the PSF positions are estimated as 2$\times$FWHM/signal-to-noise \citep{mendez2014, lin2021}. We note that \textit{HST} UVIS and IR imaging of J1721+8842 was taken as part of proposal 15652 (P.I.: T. Treu), and shows the same configuration as discovered in our deconvolved data (Schmidt et al. 2021, in prep.). In the following Sections, we characterise the nature of this system through spectroscopy and lens modelling.

\begin{table}
\caption{Measured astrometry and photometry from the deconvolution of NOT StanCam imaging of J1721+8842. Uncertainties are estimated based on the signal-to-noise as described in the text. Photometry is measured on the best-seeing single epoch (from 30 May 2021). The galaxy position and uncertainty are measured by fitting a PSF-convolved Sersic to the extended component of the image (panel (c) of Fig. \ref{deconvolution}). The magnitudes are calibrated to match those from \textit{Gaia}, and so should not be taken as an absolute calibration given the different bandpasses.}             
\label{table:astrometry}      
\centering                          
\begin{tabular}{c c c c}         
\hline\hline                 
Image & $\Delta$ R.A. ($\arcsec$) & $\Delta$ Dec. ($\arcsec$) & $r$ (mag) \\    
\hline                        
    A & -1.348 $\pm$ 0.003 & 0.966 $\pm$ 0.003 & 18.21 $\pm$ 0.02  \\      
    B & 0.054 $\pm$ 0.003 & 2.293 $\pm$ 0.003 & 18.62 $\pm$ 0.02 \\
    C & -1.227 $\pm$ 0.003 & -1.387 $\pm$ 0.003 & 19.01 $\pm$ 0.02 \\
    D & 1.915 $\pm$ 0.004 & -0.340 $\pm$ 0.004 & 19.93 $\pm$ 0.02 \\
    E & -0.747 $\pm$ 0.015 & -2.644 $\pm$ 0.015 & 20.60 $\pm$ 0.03 \\
    F & 1.35 $\pm$ 0.07 & 1.11 $\pm$ 0.07 & 22.50 $\pm$ 0.10    \\ 
    G & 0.494 $\pm$ 0.018 & 0.120 $\pm$ 0.018 & ---\\
\hline                                   
\hline   
\end{tabular}
\end{table}

\section{NOT Spectroscopy} \label{notspectrum}
\begin{figure*}
\centering
\includegraphics[width=\textwidth, trim={0 0 0 0},clip]{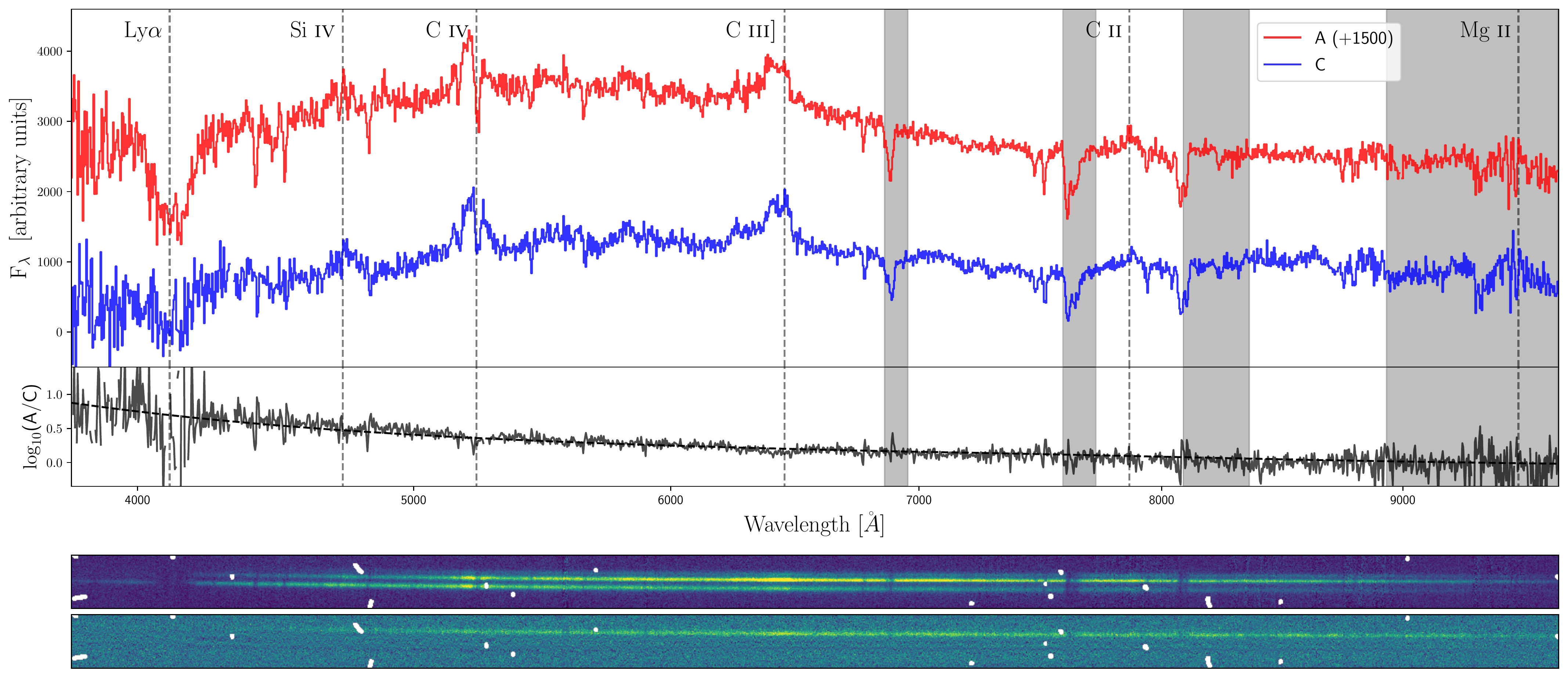}
    \caption{\textit{Top}: Spectra of A and C, and their flux ratio, with emission lines at $z=2.383$ marked. Regions with strong telluric absorption are shaded. The best-fit reddening from an SMC-like galaxy at the source redshift is overlaid on the flux ratio. \textit{Middle}: bias- and sky-subtracted two dimensional spectrum from the night of 17 April 2021. \textit{Bottom}: the same 2D spectrum after removing the traces of A and C, revealing the trace of point source E. The \textsc{C iii]} emission line is clearly visible, and the 1D spectrum can be found in Fig. \ref{spectrum2}. The spectrum of A shown in the top panel is from 16 April 2021, since the slit was unknowingly offset to the West, thus capturing more flux from A.}
          \label{spectrum}%
\end{figure*}

To determine the nature of E and F, follow-up long-slit spectroscopy was obtained with ALFOSC on the NOT. On both 16 and 17 April 2021, a 1800s spectrum was taken with grism \#4, providing a wavelength coverage of 3200--9600\,\AA\ and dispersion of 3.3\,\AA\ pixel$^{-1}$. The seeings on the two nights were $\sim$1.2 and $\sim$0.75\arcsec\ respectively. Given the faintness of image F, and its proximity to the lensing galaxy, we targeted only image E. We aligned C and E in the slit to minimise contamination, as shown in panel (a) of Fig. \ref{deconvolution}. This also ensured a reliable estimate of the 1D PSF model at all wavelengths. Given the constant airmass of the system throughout the night, we chose to observe the system when the atmospheric dispersion was in the same direction as the slit (i.e., when the parallactic angle matched the slit angle) in order to avoid the throughput loss and systematic effects of using the atmospheric dispersion corrector. A 1.0\arcsec\ slit was used. The bias-corrected, sky-subtracted, cosmic-ray-masked image from the second observation can be seen in Fig. \ref{spectrum}.

To extract separate spectra, we forward model the quasar images onto a pixelised grid, and apply a 1\arcsec-width slit. As seen in Fig. \ref{spectrum}, three traces are visible, from A, C, and E. Images C and E are partially blended. We therefore model only these three sources, using the \textit{Gaia} astrometry to place them on the 4-times-oversampled pixelised grid. We assume each component is described by a circular Moffat profile, and we neglect any flux from the lensing galaxy (the closest approach to the slit being $\sim$ 2\arcsec). Given the high airmass ($\sim 2.0$) and the slight offset between the slit angle and parallactic angle ($\sim$12 degrees), we also account for slit losses from atmospheric differential refraction. Perpendicular offsets from the slit are calculated following \citet{filippenko1982}. We further allow for a possible offset in the slit centering which we constrain by requiring the extractions from each night to have similar flux ratios, while also having clean 2D spectrum residuals.

To determine the Moffat parameters and the absolute position in the slit at each wavelength, we perform the following steps:

\begin{enumerate}
\item we bin the reduced data into 30 wavelength-bins of equal signal-to-noise, 
\item each binned profile is fit using three 1D PSF profiles (found by convolving a 1\arcsec-width rectangular slit with three circular Moffat profiles for A, C, and E, positioned based on \textit{Gaia} astrometry, and summing across the slit). The relative position in the slit and PSF parameters (FWHM and Moffat $\beta$), and their associated uncertainties, are found using MCMC sampling with the \textsc{emcee} Python package, while allowing for a constant background, and with a uniform prior on the Moffat $\beta$ parameter between 1 and 10 \citep{foremanmackey2013},
\item with the 30 measurements and uncertainties of these three parameters, we model  the wavelength changes of each using a polynomial fit of order 4,
\item at each pixel in the spectral direction, the three PSF components are generated using the PSF parameters and absolute position from the fitted polynomial functions,
\item the fluxes and uncertainties are extracted through a linear least squares fit allowing for a constant background,
\item spectra are then corrected for instrumental response using the spectrum of a standard star, and additional atmospheric extinction.
\end{enumerate}

\begin{figure*}
\centering
\includegraphics[width=\textwidth]{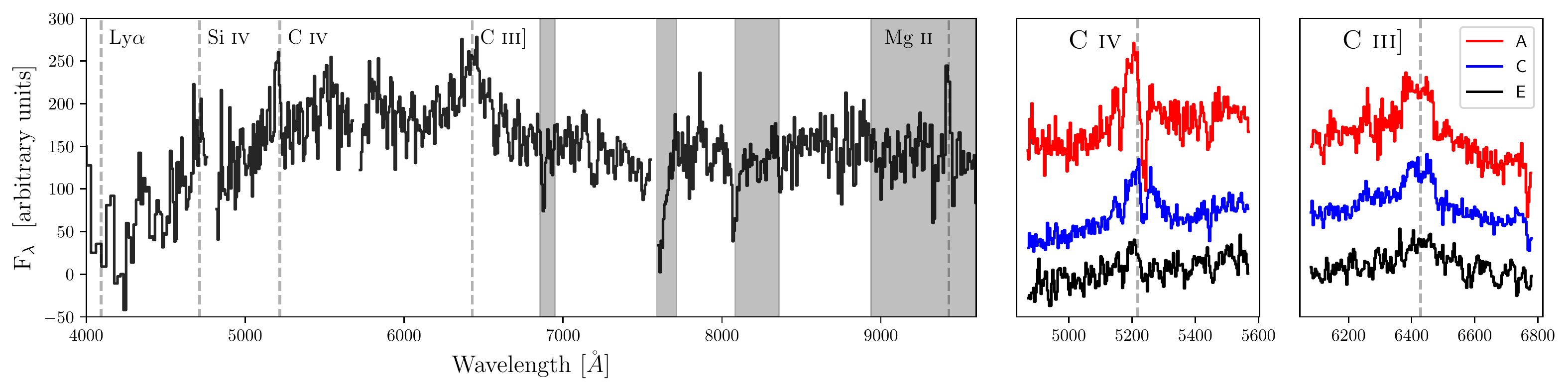}
\caption{Spectrum of E, showing broad \textsc{C iii]} emission, as well as detections of Si \textsc{iv}, C \textsc{iv}, and Mg \textsc{ii}. Regions with strong telluric absorption are shaded. Zoom-ins of the \textsc{C iv} and \textsc{C iii]} of A, C, and E are shown for comparison, with arbitrary offsets and E multiplied by 2.6.}
          \label{spectrum2}%
\end{figure*}

The extracted spectra are shown in Fig. \ref{spectrum}. The 2D spectrum with and without the two images of the quad (A and C) subtracted, is also shown, with the latter leaving just the 2D spectrum of E. The extracted spectrum of E is shown in Fig. \ref{spectrum2} with an equal-noise binning to highlight any features in the noisy blue end of the spectrum. Also shown are zoom-ins of the \textsc{C iii]} and \textsc{C iv} emission lines of the three quasar spectra. 

\subsection{The nature of E}
In the 2D spectrum of E, an emission line is seen at the same wavelength as \textsc{C iii]} for the quad images (see the bottom panel of Fig.~\ref{spectrum}). The 1D extracted spectrum of E (Fig.~\ref{spectrum2}) confirms this emission line to be \textsc{C iii]}, due to the presence of other emission lines which place E at a similar redshift to the source of the quad. We derive redshifts of each emission line by fitting a Gaussian and a linear continuum background to data within 350\,\AA\ of the expected wavelength with MCMC sampling (see Table \ref{table:redshifts}). We take the \textsc{C iii]} derived redshift as the fiducial redshift of this system, however this can differ significantly from the systemic redshift of the source quasar, due to a bias from absorption, the presence of other broad emission lines, and a systematic blueshift of broad emission lines often seen in quasars \citep[e.g.,][]{rankine2020}. We compare this redshift to that derived from the emission lines seen in the spectra of A and C. We follow the same procedure, this time masking any strong absorption, and fitting \textsc{C iii]} and Al \textsc{iii} simultaneously. Medians and 1$\sigma$ uncertainties are shown in Table \ref{table:redshifts}. Also present in the spectra of A and C is narrow Lyman-$\alpha$ (see Sect. \ref{PDLA}), possibly from the host galaxy, which we measure at slightly higher redshift. If this Lyman-$\alpha$ emission is originating from the host galaxy, then we can take it as our systemic redshift, yielding a \textsc{C iv} blueshift of $\approx 2300 \pm 600$ \kms\ in the quad source. 

The derived redshift of E based on its \textsc{C iii]} line -- $z=2.369\pm0.007$ -- is consistent with the broad emission lines of A and C, suggesting it is either another image of the quad source, or the source is a binary quasar. While the signal-to-noise of the spectrum is too low to make a useful spectral comparison, the \textsc{C iii]} and Mg \textsc{ii} profiles appear qualitatively different from those of A and C. In Sect. \ref{modelling}, we use lens modelling to show that E and F must belong to a second quasar source, offset in the source plane by 0.73\arcsec\ corresponding to 6.0 kpc at $z = 2.38$. Given the similarity of the measured redshifts of each system and very small sky separation, the source is very likely a physical binary quasar.

\begin{table}
\caption{Measured redshifts of emission and absorption lines in the NOT spectrum, shown above and below the line respectively. All uncertainties are statistical. Absolute and relative systematic uncertainties may exist at the level of $\sim$1 pixel, corresponding to 3\,\AA\, or a redshift uncertainty of 0.002 around C $ \textsc{iv}$. }             
\label{table:redshifts}      
\centering                          
\begin{tabular}{c c c c}         
\hline \hline                 
Line & $z_{\textrm{A}}$ & $z_{\textrm{C}}$ & $z_{\textrm{E}}$  \\    
\hline                            
\textrm{Ly$\alpha$} & 2.393$\pm$0.006 & 2.391$\pm$0.005 & -- \\
\textrm{Si} \textsc{iv}+\textsc{O iv} & 2.380$\pm$0.003 & 2.385$\pm$0.004 & 2.37$\pm$0.06  \\
\textsc{C iv} & 2.361$\pm$0.003 & 2.368$\pm$0.003 & 2.35$\pm$0.04 \\
\textsc{C iii]} & 2.364$\pm$0.002 & 2.364$\pm$0.002 & 2.369$\pm$0.007  \\
\textsc{C ii} & 2.383$\pm$0.003 & 2.384$\pm$0.006 & --  \\
\textrm{Mg} \textsc{ii} & 2.392$\pm$0.004 & 2.395$\pm$0.004 & 2.34$\pm$0.04  \\
\hline

O $\textsc{i}$--1335 & 2.3843$\pm$0.0005 & 2.3883$\pm$0.0016\\
C $\textsc{ii}$--1335 & 2.3843$\pm$0.0005 & 2.3879$\pm$0.005\\
Si $\textsc{ii}$--1527 & 2.3841$\pm$0.0005 & 2.3871$\pm$0.0010\\
Fe $\textsc{ii}$--1609 & 2.3851$\pm$0.0010 & 2.3868$\pm$0.0004\\
Al $\textsc{ii}$--1671 & 2.3838$\pm$0.0005 & 2.3866$\pm$0.0005\\
Si $\textsc{ii}$--1808 & 2.3835$\pm$0.002 & 2.3849$\pm$0.005\\
Fe $\textsc{ii}$--2344 & 2.3848$\pm$0.0004 & 2.3854$\pm$0.0004\\
Fe $\textsc{ii}$--2587 & 2.3830$\pm$0.0005 & 2.3844$\pm$0.0018\\
Fe $\textsc{ii}$--2600 & 2.3834$\pm$0.0009 & 2.3840$\pm$0.0008\\

\hline \hline
\end{tabular}
\end{table}

\subsection{A proximate Damped Ly$\alpha$ Absorber} \label{PDLA}
The spectra of A and C both show the signature of a proximate damped Ly$\alpha$ absorber (PDLA), namely neutral hydrogen absorption at or near the source redshift. Such systems are rare but useful tools for understanding the environments of quasars, either by studying the statistics, kinematics, and chemistry of neighbouring galaxies, or by acting as coronographs of the central broad line region, allowing more detailed study of the extended narrow \textrm{Ly$\alpha$} emission from the host galaxy \citep{cai2014, ding2020}. \citet{finley2013} searched for strong PDLAs amongst 88 000 SDSS spectra finding 31 examples, of which $\sim$25\% had narrow \textrm{Ly$\alpha$} emission, suggesting two different mechanisms for the absorption, either by \textsc{Hi} clouds within the quasar host galaxy \citep{fathivavsari2015}, or by extended neutral hydrogen in a neighbouring galaxy along the line-of-sight \citep{ellison2010}. \citet{xie2018} study the residual flux in the Ly$\alpha$ troughs of two PDLA quasars, suggesting the PDLA of one to be within the host galaxy, and the other to be caused by an infalling or nearby galaxy. \citet{zafar2011} studied the 3.3\arcsec \ separation binary quasar, Q0151+048 ($z=1.93$), which shows a PDLA in only one component with extended narrow Lyman-$\alpha$ emission physically associated to the quasars, and the PDLA being $>30$ kpc with an inflow speed of $\sim$600 \kms. The discovery and study of these systems are key to understanding the evolution of quasars, their host galaxies, and distributions of neutral and ionised gas in and around them at high-redshift \citep[e.g.,][]{fumagalli2017, ginolfi2018, arrigoni2019}.

\begin{figure*}
\centering
\includegraphics[width=1.0\textwidth]{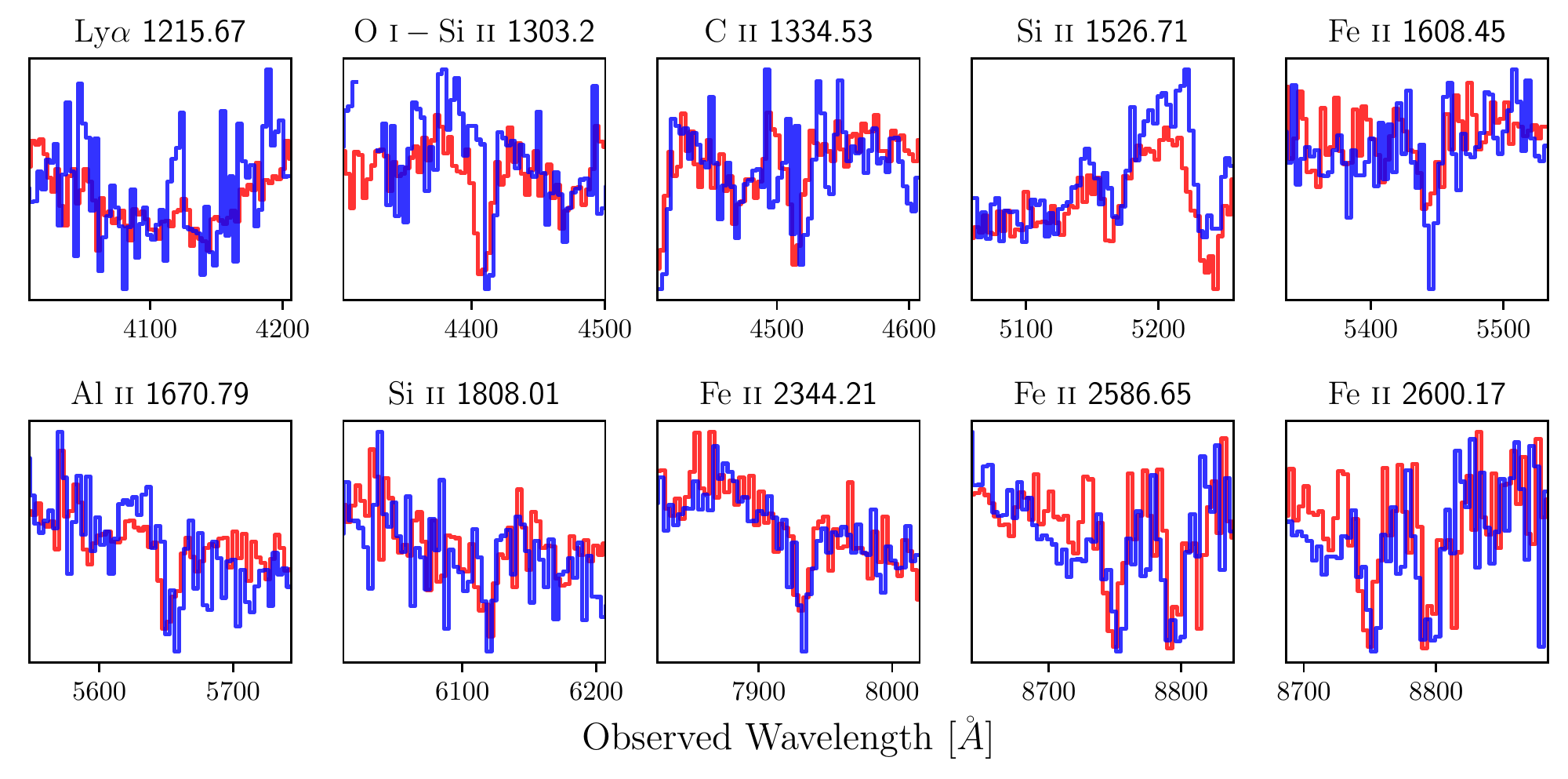}
\caption{Absorption profiles associated to the PDLA in the spectra of A (red) and C (blue). The spectra have been reddening-corrected based on the continuum, and normalised to have the same continuum flux. Redshift measurements associated to these lines are given in Table \ref{table:redshifts}. We note that C \textsc{iv} absorption is also present as seen in the Si \textsc{ii}-1526 panel. Any absorption lines falling near or within telluric absorption lines have been excluded, such as Fe \textsc{ii}-2382, which is likely to be present. As in Fig. \ref{spectrum}, the spectra are taken from separate epochs, and any absolute wavelength calibration should be taken to have possible systematics at the level of $\sim$1 pixel, i.e., $\sim$3\,\AA.}
          \label{fig:absorption_lines}%
\end{figure*}

Unfortunately the low-resolution of our spectra does not allow for a comprehensive analysis of the various column densities. However, we estimate the DLA redshift by fitting a simple Gaussian and linear background to the metal absorption lines, finding consistent redshifts of $z_{\textrm{DLA,\ A}}$ = 2.384$\pm$0.001 and $z_{\textrm{DLA,\ C}}$ = 2.386$\pm$0.002. Fig. \ref{fig:absorption_lines} shows these lines, and individual measurements are listed in Table \ref{table:redshifts}. We note a slight trend in measured redshift with the wavelength of individual absorption lines, which we likely attribute to systematics in the wavelength calibration. The median values are consistent or higher than the broad line redshifts of the source quasar, and clearly within some definitions of a DLA being proximate ($\sim$3000 \kms) \citep{ellison2010}. Higher-resolution spectra will allow precise velocity measurements of the multiple sightlines, shedding light on the ionisation state and velocity structure of the PDLA.

The gravitationally lensed nature of J1721+8842 provides a direct observational probe of unprecedentedly small physical scales at high-redshift. Assuming the PDLA is 30--300 kpc from the first quasar source --- typical values found by studying photoionisation by radiation from the nearby quasar \citep{ellison2010} --- we estimate that the physical separation of the four quad sightlines at the DLA range between 0.05 and 1 pc \citep[see Equation 1 and Fig. 3 from][]{smette1992}. These scales are an order of magnitude lower than those probed even at low redshift, typically parsec-scales \citep[e.g., ][]{biggs2016, gupta2018}. An illustration of the possible scales and structure of this system is shown in Fig. \ref{geometry}.


While the spectrum of E is low signal-to-noise, we are still able to look for signs of the PDLA. The shape of the spectrum in the blue seems to show strong absorption around Ly$\alpha$. However, systematic biases from atmospheric refraction could also explain this lack of flux. More reliable is detecting the absorption from metals associated to the DLA (see Fig. \ref{fig:absorption_lines}). Figure~\ref{spectrum2} shows a zoom-in of the C \textsc{iv} emission line for each of our spectra. In A and C, strong Si \textsc{ii} and C \textsc{iv} absorption is seen, effectively producing a narrow C \textsc{iv} emission profile. There is tentative evidence for this absorption also in E, given the narrow profile. While transverse sizes of DLAs are naturally difficult to measure given a single bright source, some studies suggest their average sizes are several kpc, supporting the possibility of seeing a PDLA signature in E as well \citep{cooke2015, matawari2016, dupuis2021}. However, deeper spectra are needed to conclusively say that the PDLA covers the second quasar source.

\subsection{Reddening} \label{reddening}

The flux ratio in Fig. \ref{spectrum} shows that C is redder than A. Given the slit position angle was chosen to target C and E, the offset of A from the slit can lead to smooth systematic biases in its spectrum, which should be kept in mind when analysing any global features like reddening. However, re-extraction of the red-arm data of the original WHT confirmation spectra from \citet{lemon2018} (with A and C well aligned in the slit) confirm the reddening of C compared to A. While contamination from the lensing galaxy in the spectrum of C could account for such an observation, we find this unlikely given the lack of lens galaxy absorption signatures in the spectrum of C, and because A is more closely aligned with the galaxy position angle (see Fig. \ref{deconvolution}). 

Correcting for Galactic extinction following \citet{schlafly2011}, we obtain a spectrum of image A compatible with a typical quasar spectrum \citep{Selsing2016}, however naturally the extra reddening of C remains. This chromaticity can originate from differential reddening and/or microlensing. While reddening will affect both broad lines and the continuum, microlensing will mostly affect the continuum emission. Indeed, the latter arises from a region comparable to the microlensing Einstein radius while the broad lines arise from a region typically 5-10 times larger. Given the small change in flux ratio between the broad line fluxes and their adjacent continuum, we first decide to explore differential dust reddening.

We consider both a Small Magellanic Cloud \citep[SMC;][]{gordon2003} and Milky-Way (MW) extinction model, placed at the lens or source redshift. We mask regions of the spectra bluer than Ly$\alpha$, or with broad lines, strong absorption, or telluric lines. The observed flux ratio variation is reproduced using an SMC-like model with $E(B-V) \in [0.13-0.15]$ \,mag at the source redshift or $E(B-V) \in [0.39-0.52]$\,mag at the lens redshift. The range of extinction values corresponds to analyses of the extracted spectra from both nights of observation. Comparison to reddening of other PDLA quasars suggests reddening at the source redshift is much more likely \citep{finley2013}, and is shown as a fit to the flux ratio in Fig. \ref{spectrum}. The fit of a MW-like model is significantly poorer in the blue despite one additional parameter in the model ($R_V$). Given the reddening-corrected spectra, we can now compare the broad emission lines and discuss microlensing. As expected, a flat ratio in the continuum compatible with 1 is found, except at the wavelengths of the broad emission lines where A/C reaches about 0.8 for the various lines. This is the signature of a small amount of differential microlensing between the continuum and the emission lines: either image A is demagnified or image C is magnified. A decomposition of the pairs of spectra using the so-called MmD technique \citep{Sluse2007, Hutsemekers2010} --- a method that linearly combines the pairs of spectra assuming that one is minimally affected by microlensing --- confirms a small amount of microlensing of the continuum of $\mu_C \sim 1.6 \pm 0.2$, and supports the absence of residual chromaticity caused by microlensing.

As PDLA quasars are observed to have redder colours than other quasars \citep[e.g.,][]{geier2019}, the reddening could be associated to the DLA. Given the likely sub-parsec transverse separations betwen the images at the distance of the DLA, this would imply the existence of compact dust structures on sub-parsec scales, e.g., analogues of Bok globules seen in the Milky Way, which has been previously observed in known quasar DLAs \citep{krogager2016, bergeron2017}. However, microlensing or dust on larger scales more distant from the quasar cannot be ruled out. Deeper data, especially at blue wavelengths and for all 6 quasar images, will be crucial in distinguishing these scenarios.

\section{WHT Lens Spectroscopy} \label{whtspectrum}
\begin{figure*}
\centering
\includegraphics[width=\textwidth]{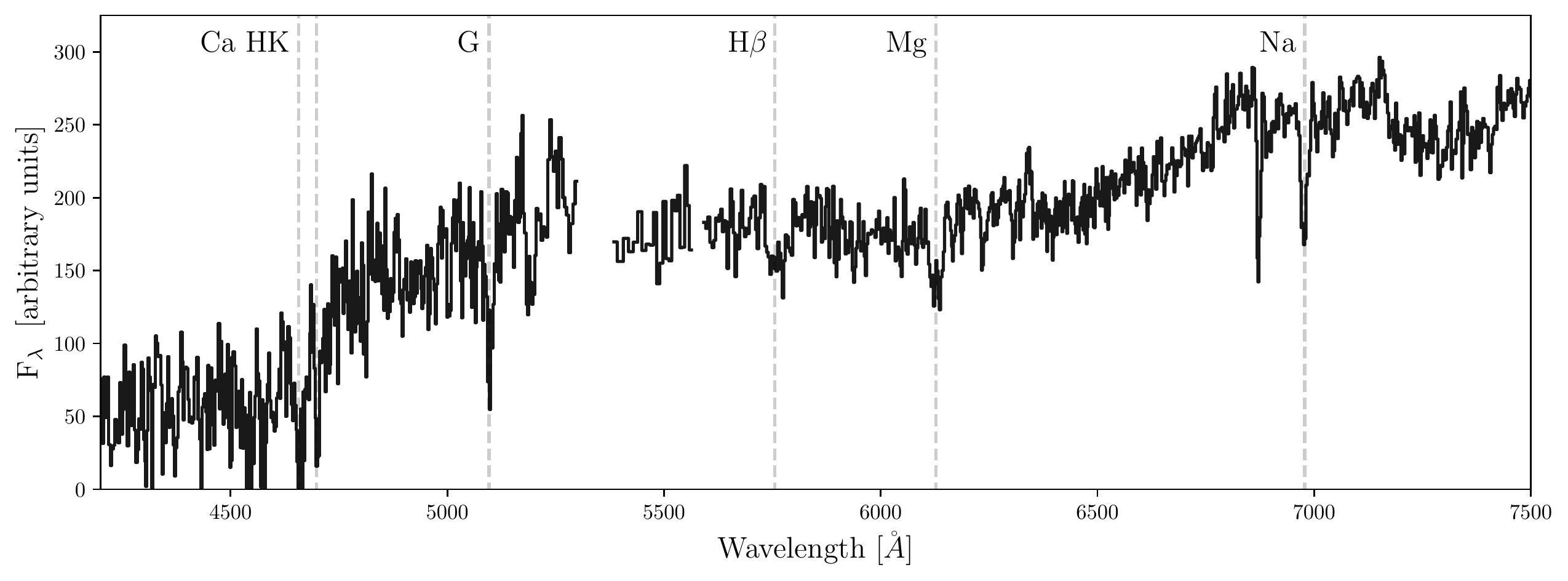}
\caption{WHT-ISIS spectrum of the lensing galaxy in J1721+8842. There are clear absorption lines associated to a luminous red galaxy at $z=0.184$. The relative flux calibration between the red and blue arms is not absolute, and the noisiest part of the spectrum (between 5300 and 5380\,\AA) has been omitted. }
          \label{whtspec}%
\end{figure*}

Long-slit spectra targeting the redshift of the lensing galaxy were taken with the Intermediate-dispersion Spectrograph and Imaging System (ISIS) on the William Herschel Telescope (WHT) on 2 August 2018 with grisms R316R and R300B for the red and blue arms respectively. Six 1000s exposures were taken with the position angle 95 degrees East of North, chosen to be aligned along the best-fit galaxy position angle based on the original Pan-STARRS $r$-band data. We note that the parallactic angle varied from 94 to 68 degrees during the observations. The observations were reduced with the same methodology as for the NOT spectra, fitting three PSFs as images B and D show significant flux in the slit. Due to atmospheric refraction, and the lack of corrector on the WHT, blue flux was evidently lost below 4500\,\AA. The separate epochs were combined after flux-matching and interpolating to a common wavelength basis. The equal-noise-binned galaxy spectrum is shown in Fig. \ref{whtspec}. 

There are clear absorption lines from Ca H and K, G-band, Mg, Na, and H$\beta$. Fitting each with a Gaussian profile provides a consistent redshift estimation of $z=0.1841\pm0.0005$. We note that this is a particularly low redshift compared to the lensing galaxies of known lensed quasars. For comparison, only two published lens redshifts are lower: SDSSJ1155+6346 at $z=0.176$ \citep{pindor2004}, and Q2237+030 at $z=0.039$ \citep{huchra1985}. 


\section{Lensed Binary or Six-image Lens?} \label{modelling}
The immediate configuration of point sources is suggestive of a quad and a double, however, without better quality spectra of E and F, we cannot immediately rule out a common source for all 6 images. Though lenses with more than four point-like images are rare, several examples already exist within the currently-known list of $\sim$200 lensed quasars. These are either due to multiple lensing galaxies \citep[e.g., B1359+154,][]{rusin2001}, or multiple compact sources detected thanks to high-resolution radio data \citep[i.e., B1933+503,][]{sykes1998}. To establish the novel configuration of J1721+8842, we consider two hypotheses: (i) the images are all from one source, or (ii) from multiple sources. While purely elliptical smooth potentials can only give rise to a maximum of 4 images, the addition of a shear perpendicular to the lens ellipticity can cause `swallowtails' to form and overlap in the astroid caustic, leading to source regions forming 6 or 8 images \citep{witt2000}. \citet{keeton2000} describes this full family of models, as well as two-galaxy models, showing the latter to be a more likely candidate for observing 6-image systems, given the former's strong requirement on shear. \citet{evans2001} argue that a more likely producer of 6 and 8 image lenses is by way of embedded discs or boxiness, and estimate $\sim$1\% of all lenses to comprise 6- or 8-image systems. A common feature of all these sextuplet systems is that the images lie close to a circle, which is not the case for J1721+8842. \citet{shin2008} study lensing by two isothermal spheres, and find that 7-image lenses exist, but require two images to be strongly demagnified, in agreement with \citet{kochanek1988}. Given this, and the lack of other structures in the deconvolved background (Fig. \ref{deconvolution}), we limit our investigation to a single lensing galaxy. 

We choose to test Singular Isothermal Ellipsoid (SIE) and Power-law Elliptical Mass Distribution (PEMD) profiles with and without shear, as such models have shown to be effective at recovering image positions of known lenses \citep[e.g.,][]{shajib2019}. For each model/hypothesis combination, we perform both a source plane and image plane optimisation, using \textsc{lensmodel} \citep{keeton2001b, keeton2001a}, excluding the image-plane case of no-shear/single source since this can never produce 6 images. Since flux ratios can be strongly affected by microlensing, variability over the time delay, extinction, and substructures, we choose not to use them as constraints. We use only the image and galaxy positions to constrain our models, as listed in Table \ref{table:astrometry}. Similarly to the images, the likelihood contains a term constraining the lens mass position to that of the measured light position, weighted by its uncertainty.

Results are summarised in Table \ref{table:massmodels}. In all cases the two-source hypothesis is favoured, as expected given the configuration and general rules from single-source 6-image lens studies. The SIE+shear fit to two point sources has the smallest reduced ${\chi}^{2}$, thus we take this as our fiducial model. The median and 1$\sigma$ model parameter values are: Einstein radius, $b=1.977\pm0.002$\arcsec, galaxy mass position angle, $\theta_{\textrm{SIE}}=78.0\pm1.6^{\circ}$ (East of North) and axis ratio, $q_{\textrm{SIE}}=0.87\pm0.02$, the external shear, $\gamma=0.098\pm0.004$, and its position angle $\theta_{\gamma}=66.2\pm0.6^{\circ}$ (East of North). The lens galaxy mass position angle and ellipticity are in good agreement with that of the measured light, which gives $\theta_{\textrm{light}}\sim70^{\circ}$ and $q_{\textrm{light}}\sim0.81$. These small offsets are compatible with modelling analyses of known quads with multi-band \textit{HST} data \citep{shajib2019}. Best-fit estimates of the convergence, shear, and magnification at each image position are given in Table \ref{table:kappagamma}. The source separation is $0.730\pm0.014$\arcsec, corresponding to $5.95\pm0.11$ kpc at $z=2.38$. Using the model image positions, we note that the total magnifications of the quad and double are 19.6 and 6.4 respectively, and the intrinsic flux ratio of the quad source to double source is 4.4.

   \begin{figure}
   \centering
   \includegraphics[width=\columnwidth]{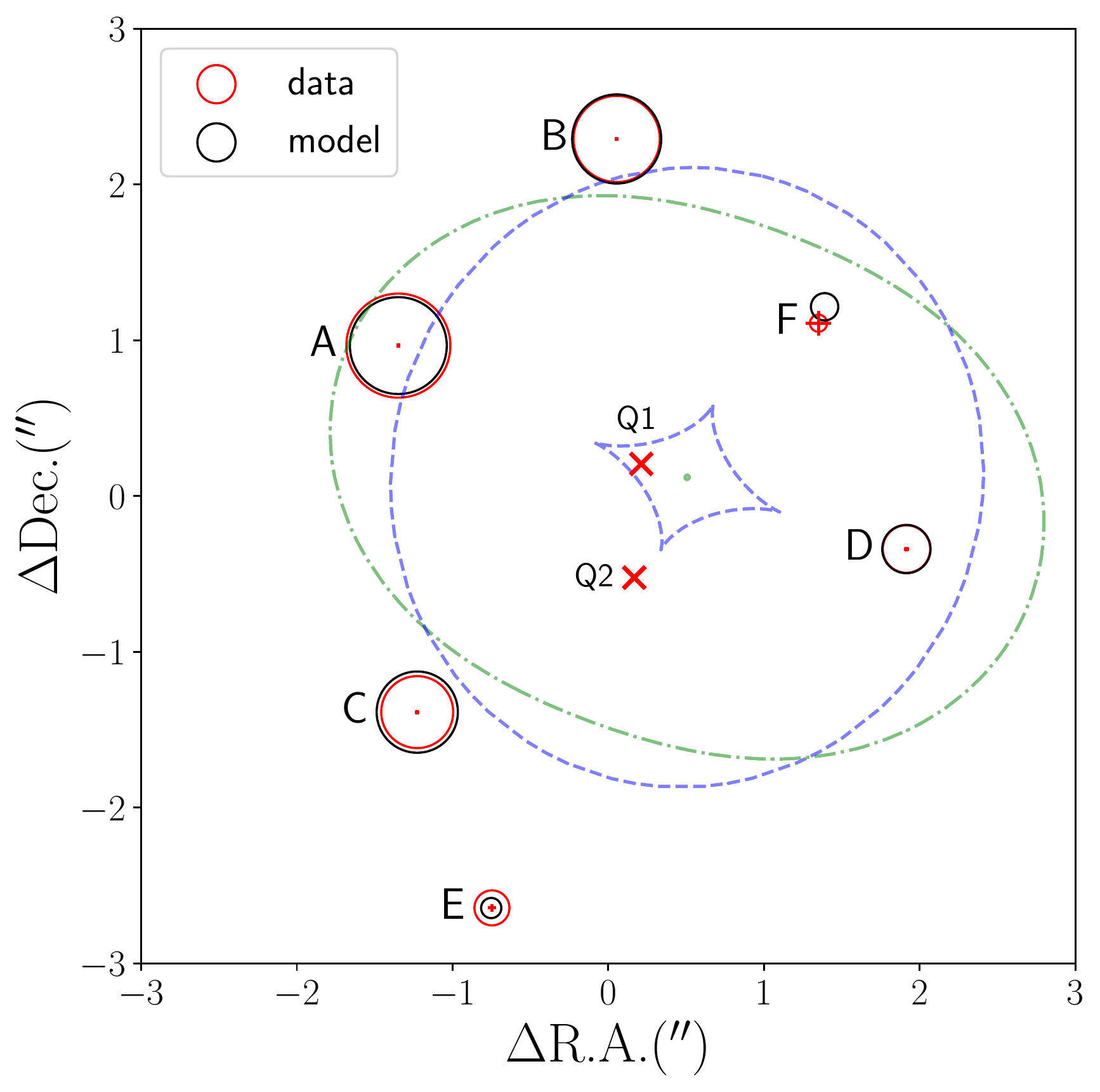}
   \caption{A summary of the SIE+shear mass model with two sources. The flux of the measured images (red), and predicted images (black), are proportional to the circle areas; these circles are centred on the relevant positions, along with red crosses showing the positions and astrometric uncertainties of the measurements. Critical curves and caustics are shown in green and blue respectively. The two quasars are marked with red crosses, with Q1 corresponding to images ABCD, and Q2 to EF.}
              \label{model}%
    \end{figure}

\begin{table*}
\caption{A summary of the various mass models fit to the point sources of J1721+8842. For the models fitting only ABCD, we cast E back into the source plane, and find any further images in the image plane (always one extra), and provide this image's distance from the measured position of F. $\gamma^{\prime}$ is the power-law slope ($\rho \ \propto r^{-\gamma^{\prime}}$).} 
\label{table:massmodels}      
\centering                          
\begin{tabular}{c c c c c c c}         
\hline\hline                 
    Mass model & Images fit & $\textrm{N}_{\textrm{sources}}$& offset from F (\arcsec)  & $\chi^{2}_{\nu} \equiv \chi^{2}$ / $\textrm{d.o.f.}$ & $\gamma^{\prime}$ & source $\Delta R$ (\arcsec)\\    
\hline                        
    SIE & ABCD & 1 & 0.68 & 50 $\equiv$ 151 / 3  & --- & --- \\
    SIE + shear & ABCD & 1 & 0.15 & 0 $\equiv$ 0 / 1& ---& ---\\
    SIE & ABCDEF & 2 & --- & 50 $\equiv$ 251 / 5  & --- & 0.88$\pm$0.01   \\
    SIE + shear & ABCDEF & 2 & --- & 1.1 $\equiv$ 3.2 / 3 & --- & 0.73$\pm$0.01 \\
    PEMD + shear & ABCDEF & 2  & --- & 1.3 $\equiv$ 2.7 / 2 & 1.84$\pm$0.06 & 0.64$\pm$0.04 \\    
    SIE + shear & ABCDEF & 1 & --- & 778 $\equiv$ 3890 / 5   & --- & --- \\

\hline
\hline
\end{tabular}
\end{table*}

\begin{table}
\caption{Local convergence, shear, and magnification values at the positions of the measured images using the best-fit SIE+shear model. Negative and positive magnifications represent saddle points and minima in the time delay surface respectively. } %
\label{table:kappagamma}      
\centering                          
\begin{tabular}{ccccc}         
\hline\hline                 
    Image & $\kappa$ & $\gamma_{1}$ & $\gamma_{2}$ & $\mu$\\
\hline                        
A & 0.516 & $-$0.403 & 0.462 & $-$7.02 \\
B & 0.424 & 0.324 & 0.241 & 5.93 \\
C & 0.420 & $-$0.123 & $-$0.344 & 4.93 \\
D & 0.713 & $-$0.641 & 0.493 & $-$1.74 \\
E & 0.305 & 0.135 & $-$0.157 & 2.28 \\
F & 0.729 & 0.049 & $-$0.647 & $-$2.87 \\
\hline
\end{tabular}
\end{table}

Since we have measured both the lens and source redshifts, we are able to predict the following time delays for the system: $\Delta t_{\textrm{AB}}=-2.5$, $\Delta t_{\textrm{AC}}=-4.6$, and $\Delta t_{\textrm{AD}}=26.0$ days for the quad, and $\Delta t_{\textrm{EF}}=79.2$ days for the double. Negative delays imply that any source variations are seen in the second component first. Another prediction from this mass model is image flux ratios. For the quad, these ratios are compatible with our single-epoch observed flux ratios within $\sim$20\%, with the main differences being A observed 15\% too bright, and C observed 20\% too faint. This missing flux in C supports our conclusion that C is being reddened and/or demagnified by microlensing in Sect. \ref{reddening}. The flux ratio of EF is very discrepant, with a measured ratio of 3.6, but a predicted ratio of 0.55 (see Fig. \ref{model}). Since E is seen in the original Pan-STARRS discovery image (from $\sim 2012$), whereas F is not, this is unlikely to be attributed to variability over the time delay. More likely is a combination of extinction, millilensing, and microlensing. The slight offset of F from the best-fit model position supports the millilensing explanation, however the uncertainty on the position is large. Deeper spectra, imaging, and photometric monitoring will address these discrepancies.

\section{Discussion} \label{discussion}
A qualitative picture of quasar activity triggered through mergers has been long-understood through simple dynamical friction models \citep{bahcall1997, kochanek1999, mortlock1999, djorgovski2007, vignali2018}, but a complete model matching observed statistics at all separations is lacking, with some studies expecting binary quasars to be a factor of 10 more frequent than observed \citep{junkkarinen2001}. Merger acceleration, obscuration, fuelling dynamics, galactic structure, and observational selection effects all likely play a role \citep{foreman2009, capelo2017, li2021}. The discovery that the true source of J1721+8842 is a 6 kpc projected-separation binary active galactic nucleus (AGN) at $z=2.38$ presents important constraints for merger models and their relation to quasar fuelling. The 3D separation is not yet constrained, as our low-resolution spectra and lack of narrow emission lines do not allow for a more precise determination of the two source redshifts. While the redshifts are consistent within the uncertainties, the data are also compatible with offsets of up to $\Delta z \sim 0.005$ (i.e., several Mpc), however the chances of observing such a system at pericentre approach in an inclined orbit for line-of-sight separations far above the transverse separation is unlikely. 

Sub-10-kpc binary AGN have been found at low-redshift, such as LBQS0103$-$2753, a 2-kpc-separation binary at $z=0.86$, showing no signs of obscuration \citep{shields2012}. Few such binaries are known to exist during `cosmic noon' -- the peak of quasar activity. \citet{chen2021} provide a compilation of known binary quasars and candidates, with only one sub-10-kpc confirmed system above $z=1$: SDSSJ1238+0105, a BAL+non-BAL quasar pair at $z=3.13$ separated by 7.8 kpc, confirmed by \citet{tang2021} (note that SDSSJ0818+0601 from \citet{more2016} has subsequently been confirmed as a lensed quasar by \citet{hutsemekers2020}, and J2057+0217 from \citet{lemon2018} is potentially also a lensed quasar). Two further systems from the literature are likely real binary quasars due to their differing spectra: DESJ0120$-$4354 \citep[7.2 kpc at $z=1.91$,][]{anguita2018}, and SDSS J1008+0351 \citep[9.4 kpc at $z=1.74$,][]{inada2008}. Hydrodynamical cosmological simulations matching the small-scale clustering of quasars above 25 kpc separation predict an increasing correlation function with decreasing separation down to their resolution of 10 kpc across all redshifts \citep{degraf2017, bhowmick2019}. Extrapolating this below their measured range gives a conservative estimate of 0.5 systems in 7600 square degrees from $z=0.43$ to $z=2.26$ between 0-10 kpc, with each component having $g<20.85$  \citep{eftekharzadeh2017}. The unlensed magnitudes of our system are estimated to be $\sim19.9$ and $\sim22.0$, and thus represent a fainter sample. However, the discovery of this system is entirely due to the fact that it is lensed. Given that current search methods are unable to discover and confirm such small separation binaries when they are not lensed, we can take this discovery to estimate very crudely the underlying population by simply determining the chances of lensing. This has been constrained from statistical lens samples in the optical and radio, and is $\sim 10^{-3}$ \citep{browne2003, inada2008, om10}. Therefore, the expected number of unlensed systems with similar properties to the source in J1721+8842, is $\sim$ 1000, in a similar area probed by that of lensed quasar searches. Furthermore, the population of lensed quasars can be used to support this, as they represent an unbiased sampling of quasar binaries at high-redshift. Assuming that all lensed quasars are sufficiently well-studied to identify any binary sources lying entirely within their $\sim$ 1\arcsec-radius multiple-imaging source region, we can constrain the rate of binaries at high-redshift to be the number of known lensed binaries divided by the number of known lensed (non-binary) quasars. Approximately 200 lensed quasars are known \citep{lemon2019}, suggesting a binary fraction rate of 0.5\% below $\sim$10 kpc. \citet{schwartz2021} have found evidence for a sub-kpc binary within another known lensed quasar, suggesting an even larger binary fraction rate. Given $\sim 200,000$ spectroscopic quasars from SDSS above $z=2$ \citep{paris2018}, we then expect $\sim$1000 sub-10 kpc binary quasars to already exist at the photometric limit of SDSS spectroscopy, namely $r\sim22$. Selection effects for spectroscopic targeting, namely requiring morphological consistency with a PSF, likely prevent these systems from having archival spectra. The ability to discover such systems has thus been hindered, and made worse by the possible large flux ratios between the components, however these binary systems should still exist within the SDSS footprint. While these arguments rely on just a single detection, and neglect important considerations of binary flux ratios and separations, caustic area distributions, and details of lensed quasar follow-up, it is nonetheless strongly suggestive of a large number of undiscovered binary quasars at small separation above $z=2$. There is precedent for such enhanced late-stage fuelling and dual AGN fraction from higher resolution simulations and observations \citep{vanwassenhove2012, stemo2020}. Furthermore, recent targeted searches for small-separation quasars suggest dual AGN fractions of $\sim$0.26$\pm$0.18\% at 5-30 kpc in rough agreement with our basic estimate from lensed quasar statistics \citep{silverman2020}. 

J1721+8842 is also remarkable in the fact that it has a proximate damped Lyman-$\alpha$ system. High H\textsc{i}-column density PDLAs (\textrm{log}($N_{\textrm{H} \textsc{i}}) > 21.3 \ \textrm{cm}^{-2}$) are intrinsically rare; $\sim$ 1 in 3000 quasars have such systems within 1500 \kms \citep{finley2013}, despite the observed enhancement relative to an extrapolated line-of-sight prediction, due to typical over-densities in the environments of quasars \citep{ellison2010}. It is tempting to link the absorber to the binary nature of this system as it has been seen in wider-separation binary quasars \citep{zafar2011}. On the other hand, PDLAs have been seen in quasars without obvious signs of mergers, but still with possible inflowing gas at scales from 1-50 kpc \citep{law2018}. Such large scale gas streams have also been associated to mergers and quasar fuelling both observationally at lower redshift \citep{johnson2018}, and also in simulations, instigated by the strong tidal torquing of merging \citep{kazantzidis2005}. Whether the PDLA is associated simply to another galaxy in a possible group hosting both quasars, or due to gas movements brought about by the merging process, requires further data. 

Residual narrow Ly$\alpha$ flux is seen in the spectra we obtained for two images of the quad. This is a known phenomenon in which the PDLA acts as a coronograph, absorbing all Ly$\alpha$ emission from the broad line region of the quasar, but not that of the host galaxy or the local environment \citep[e.g.,][]{jiang2016}. Neglecting possible systematics from extraction, the narrow Ly$\alpha$ flux ratio is inconsistent with the continuum flux ratio. This may be explained by one or a combination of the following: (i) there is substantial microlensing and/or reddening affecting the compact continuum emission, but not the extended Lyman-$\alpha$ emission, which can then be trusted as the true image flux ratio; (ii) closely separated regions in the source plane can have large magnification differences due to the non-linear nature of lensing, thus if the emission is not spatially coincident with the quasars, it can lead to differences in flux ratio compared to the continuum emission \citep{borisova2016}. The latter hypothesis should be confirmed or rejected through spectra with much higher spatial resolution.

   \begin{figure}
   \centering
   \includegraphics[width=\columnwidth]{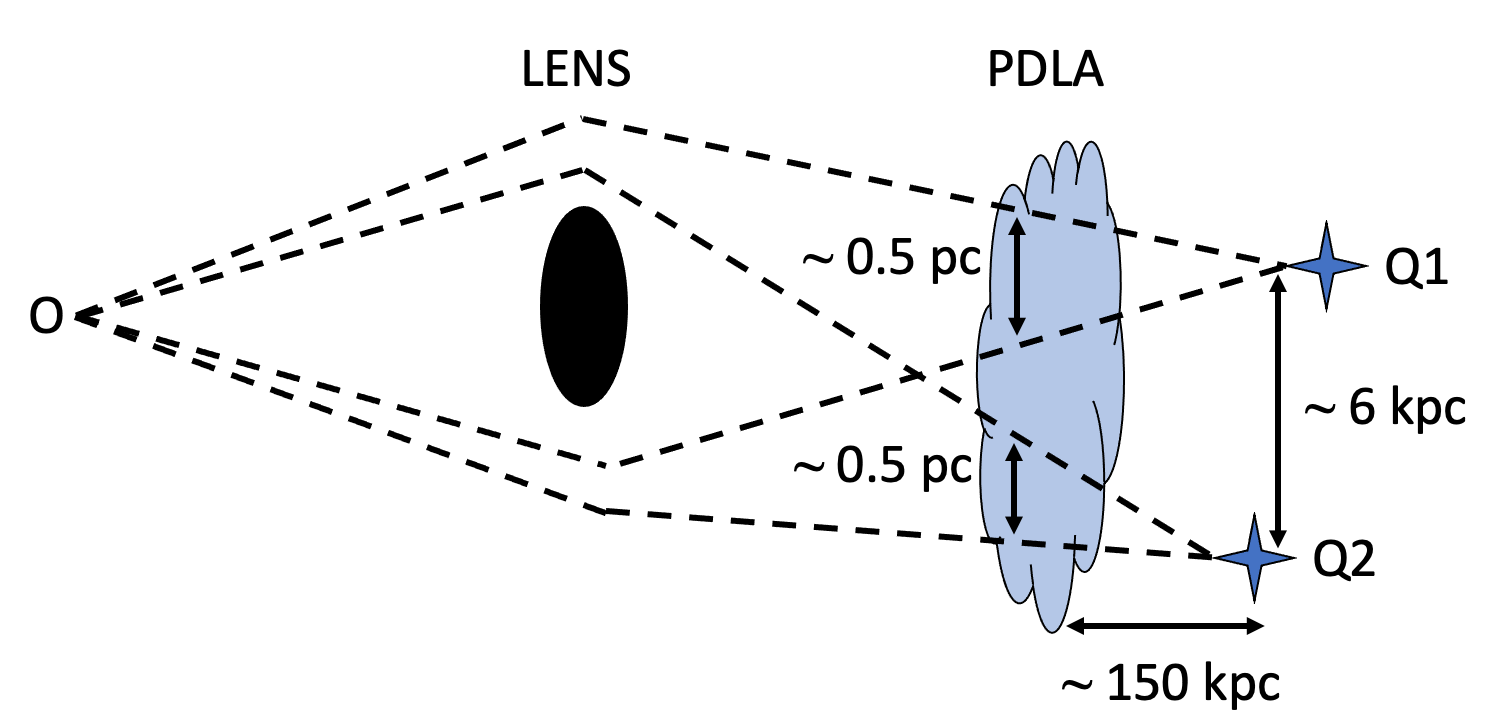}
   \caption{A sketch of the possible geometry and physical scales of the system. While we do not constrain the distance to the PDLA, typical values of known systems are in the range of tens to hundreds of kiloparsecs. This would imply a distance between common sightlines to each quasar source of $\sim$0.5 pc, and 6 kpc for sightlines between each source. We note that the PDLA might not necessarily cover both quasars, since our spectra are not deep enough to confirm this.}
              \label{geometry}
    \end{figure}

\section{Conclusions} \label{conclusions}
In this paper, we have presented a detailed follow-up in imaging and spectroscopy of the known gravitationally lensed quasar, J1721+8842. We summarise our main findings:

   \begin{itemize}
       \item Deconvolution of StanCam imaging data from the NOT reveals the presence of two new point sources either side of the lensing galaxy in J1721+8842. A long-slit spectrum shows that one is a quasar with similar redshift to the known quasar source. Lens modelling suggests these two new sources belong to a separate quasar, offset by 6 kpc from the original quasar source of the quad images, however we lack the spectral resolution to currently constrain velocity offsets.
       \item This is the fourth sub-10-kpc binary quasar above $z=1$ and smallest-separation confirmed optical binary, and would likely be undiscovered were it not for the resolving ability of lensing. The rarity of lensing suggests there should exist a large number of unlensed high-redshift small-separation quasar pairs. This is supported by the small number of known lensed quasars, which represent an unbiased probe of small-scale details in quasar source populations. We crudely estimate that $\sim$1000 sub-10 kpc binaries should exist within the SDSS footprint down to $r\sim22$. Ongoing and future lens and binary searches with surveys like HSC, \textit{Gaia}, and \textit{Euclid}, are likely to reveal many of these systems.
       \item We also analyse spectra of two images of the quad, which show a damped Lyman-$\alpha$ system. Measurements of the redshift suggest this is a proximate absorber, and thus the configuration of the system with 6 sightlines will allow simultaneous probes of this absorber at sub-parsec and kiloparsec scales. We further find evidence for narrow Lyman-$\alpha$ emission in the trough of the PDLA, which acts as a coronograph of the central quasar emission, providing a novel probe for microlensing and quasar host galaxy studies. Deeper, higher-resolution spectra are required for detailed studies, and to confirm the signs of a PDLA in the spectrum of the second quasar source.
   \end{itemize}

In brief, J1721+8842 turns out to be unique in many respects. We have shown that the source is not only a very rare close-separation binary quasar, but also hosts a rare PDLA. The unique geometry of lensing provides an invaluable tool for understanding both binary quasars and PDLAs. In addition, the system is currently monitored at the NOT to obtain time delay measurements for time-delay cosmography. Both science cases will hugely benefit from spatially resolved spectroscopy with \textit{JWST}.

\begin{acknowledgements}
We would like to thank Aymeric Galan and Giorgos Vernardos for helpful discussions. This work is supported by the Swiss National Science Foundation (SNSF) and by the European Research Council (ERC) under the European Union's Horizon 2020 research and innovation program (COSMICLENS: grant agreement No 787886). AA's work is funded by Villum Experiment Grant \textit{Cosmic Beacons} (project number 36225). This work is based on observations made with the Nordic Optical Telescope, owned in collaboration by the University of Turku and Aarhus University, and operated jointly by Aarhus University, the University of Turku and the University of Oslo, representing Denmark, Finland and Norway, the University of Iceland and Stockholm University at the Observatorio del Roque de los Muchachos, La Palma, Spain, of the Instituto de Astrofisica de Canarias. Further data were obtained through an Agreement between Aarhus University and the Ecole Polytechnique Fédérale de Lausanne (EPFL), as part of the High-cadence Lens Monitoring for Time Delay Cosmography Program. The data presented here were obtained [in part] with ALFOSC, which is provided by the Instituto de Astrofisica de Andalucia (IAA) under a joint agreement with the University of Copenhagen and NOT.
\end{acknowledgements}

%
%
\bibliographystyle{aa} 
\bibliography{references} 

\end{document}